\documentclass[twocolumn,secnumarabic,amssymb, nobibnotes, aps, prd]{revtex4-1}

\usepackage{graphicx}
\usepackage[caption=false]{subfig}
\usepackage{ragged2e}

\begin{document}

\captionsetup{justification=justified}
\captionsetup{singlelinecheck=off}

\title{Observation of non-sinusoidal current-phase relation in graphene Josephson junctions}%

\author{C. D. English, D. R. Hamilton, C. Chialvo, I. C. Moraru, N. Mason, and D. J. Van Harlingen}%
\affiliation{Department of Physics and Materials Research Laboratory, University of Illinois at Urbana-Champaign, Urbana IL 61801, USA }
\date{August 28, 2016}%
\begin{abstract}
	The current-phase relation of a Josephson junction can reveal valuable information about the processes influencing the supercurrent.  In this Letter, we present direct measurements of the current phase relation for Josephson junctions having a graphene barrier, obtained by a phase-sensitive SQUID interferometry technique.  We find that the current phase relation is forward skewed with respect to the commonly observed sinusoidal behavior for short junctions in the quasi-ballistic transport regime, consistent with predictions for the behavior of Dirac fermions in a Josephson junction.  The skewness increases with critical current and decreases sharply with increasing temperature.  
\end{abstract}
\maketitle

The interplay of superconductivity and the unique electronic structure of graphene leads to unusual coherence effects such as gate-tunable supercurrents \cite{heerche} and specular Andreev reflection \cite{b_old}. Much recent work has focused on superconductor-graphene-superconductor (S-g-S) Josephson junctions in which theoretical \cite{gonzalez,feigelman} and experimental studies \cite{shailos,miao,andrei1,Mizuno,GilhoLee} have examined the effects of parameters such as the junction geometry and barrier thickness on the critical current. However, unique information about the processes influencing the supercurrent can be obtained by measuring not just the magnitude of the supercurrent but also its dependence on the phase difference across the junction, characterized by the Josephson current-phase relation (CPR). The simplest models of Josephson tunneling predict a sinusoidal variation of the current with phase. However, deviations such as skewness are known to occur in unique systems like point contacts \cite{koops, rocca} and metallic junctions \cite{fuechsle, likharev}. It is possible to extract some information about the CPR of a junction by measuring critical current diffraction patterns, Shapiro steps, or switching current in a SQUID (superconducting quantum interference device) configuration \cite{girit}, but the most definitive approach is to measure the CPR directly using a phase-sensitive interferometry technique.  

In this Letter, we present experimental measurements of the CPR in Josephson junctions having a single-layer graphene barrier. The junction is incorporated into a superconducting loop coupled to a dc SQUID which allows the junction phase to be extracted directly (see Fig.1a). We observe significant deviations from the typical sinusoidal behavior for short S-g-S junctions, where the junction length (\textit{L}) is less than the superconducting coherence length ($\xi$) in the junction. The deviations consist of a forward (positive) skewness in the CPR that varies as a function of critical current and is detectable at temperatures below 300mK. While this behavior  is similar to that of a disordered superconductor-normal-superconductor (SNS) junction, it also appears to be well-described by self-consistent tight-binding Bogoliubov de-Gennes (TB-BdG) calculations which consider the Dirac spectrum of graphene \cite{ABSprimary}. The CPR curves of junctions with \begin{math} L > \xi \end{math} do not exhibit significant skewness. 

\begin{figure}%
\subfloat[][Interferometer Setup]{\includegraphics[width=6cm]{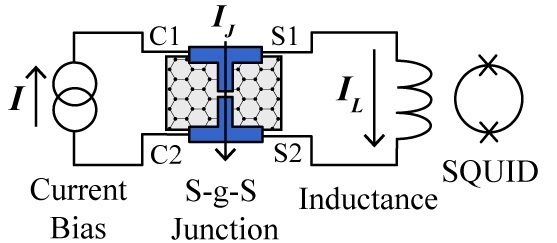}}%
\subfloat[][S-g-S Junction]{\includegraphics[width=2cm]{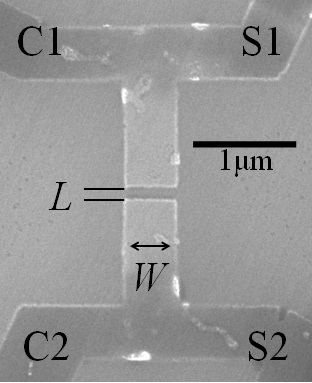}}%

\caption[]{(a) Circuit diagram of the current-phase interferometry experiment.  The graphene is depicted by the honeycomb lattice and the junction is shown on top. Note that the inductance is coupled to the SQUID through a flux transformer (not shown). (b) SEM image of the S-g-S junction. Labels S1, S2, C1, C2 correspond to identical labels in figure (a).}

\label{fig:cont1}
\end{figure}

\begin{table}
\begin{tabular}{|c|c|c|c|c|c|}
  \hline
  Sample & $L$(nm) & $W$($\mu$m) & ~$L/\xi$~ & $I_{co}$(nA) & $J_{co}$(nA/$\mu$m) \\
  \hline
  A & 70 & 0.3 & 0.35 & 71 & 237 \\
  \hline
  B & 100 & 0.5 & 0.5 & 107 & 213 \\
  \hline
  C & 350 & 3 & 1.75 & 39 & 13 \\
  \hline
  D & 350 & 10 & 1.75 & 160 & 16 \\
  \hline
  \end{tabular}
  \caption{Table summarizing the characteristics of each junction. $I_{co}$ and $J_{co}$ are the critical current and critical current density ($I_{co}/W$) of the junctions at the Dirac point at low temperature ($T<50$mK).}
\end{table}

Measurements were performed on four junctions on four separate graphene flakes. The sample dimensions and characteristics are given in Table I. All samples were prepared by mechanical exfoliation of graphite flakes onto highly $p$-doped Si substrates covered by 300 nm of SiO$_{2}$, where the doped substrate acts as a global backgate. Samples were annealed at a temperature of 400 C in 1900/1700 sccm H$_{2}$/Ar, and subsequently characterized by optical imaging, atomic force microscopy, and Raman spectroscopy. Ti(4nm)/Al(60nm) contacts were fabricated via electron beam lithography and electron beam evaporation. Samples A and B were fabricated on large flakes of graphene, with the area of the graphene much larger than the junction size (see Fig. 1b).  Junctions C and D were fabricated on narrow strips of graphene, with the edges of the graphene naturally defining the width of the junctions. Measurements were performed in a dilution refrigerator having a base temperature of 10 mK. 

Transport properties such as the mobility ($\mu$) and the residual impurity doping ($n_{o}$) are estimated from \begin{math} n = (1/2)(n_{G}+\sqrt{n_{G}^{2}+4n_{o}^{2}}) \end{math} and  \begin{math} R_{SH}=(qn\mu)^{-1} \end{math} where $q$ is the electric charge, $n_{G}$ is the carrier density electrostatically induced by the gate voltage ($V_{G}$), $R_{SH}$ is the graphene sheet resistance, and $n$ is the total carrier concentration \cite{dorgan, kim}. Using these equations to fit to the measured $R$ vs $V_{G}$ curves, we extract, on average, $\mu$ $\sim 3500$ cm$^{2}$V$^{-1}$s$^{-1}$ at $n$ $\sim$ 5 x 10$^{12}$ cm$^{-2}$, and $n_{o}$ $\sim$ 5 x 10$^{11}$ cm$^{-2}$. On SiO$_{2}$, the impurity concentration $n_{o}$ typically reduces the mean free path $l$ to 30 - 100 nm \cite{andrei2, martin}. Thus, samples A and B, where the junction length $L$ is of order $l$, are considered to be in the quasi-ballistic regime, whereas samples C and D, where \begin{math} L \gg l \end{math}, are considered to be in the purely diffusive regime. The superconducting coherence length in the junction is estimated to be \begin{math} \xi \sim \sqrt{\hbar D / \Delta} \sim 200 \end{math} nm, where \textit{D} is the diffusion length, and $\Delta$ is the superconducting gap of the electrodes \cite{tinkham}. We can then characterize samples A and B, where \begin{math} L < \xi \end{math}, as in the short junction limit, and samples C and D, where \begin{math} L > \xi \end{math}, as in the long junction limit.

\begin{figure}
\subfloat[][Short Junctions]{\includegraphics[width=7.4cm]{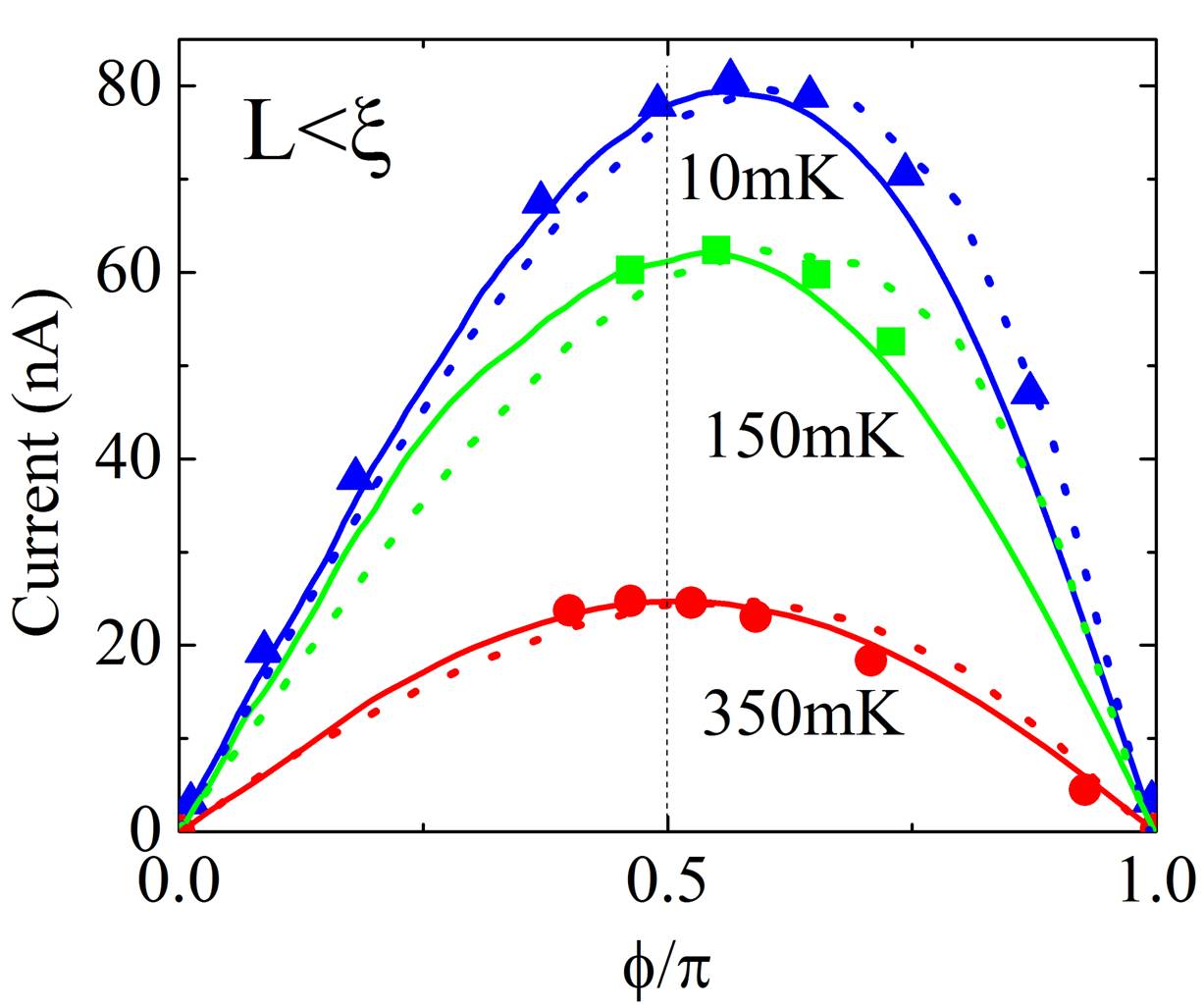}}
\qquad
\subfloat[][Long Junctions]{\includegraphics[width=7.4cm]{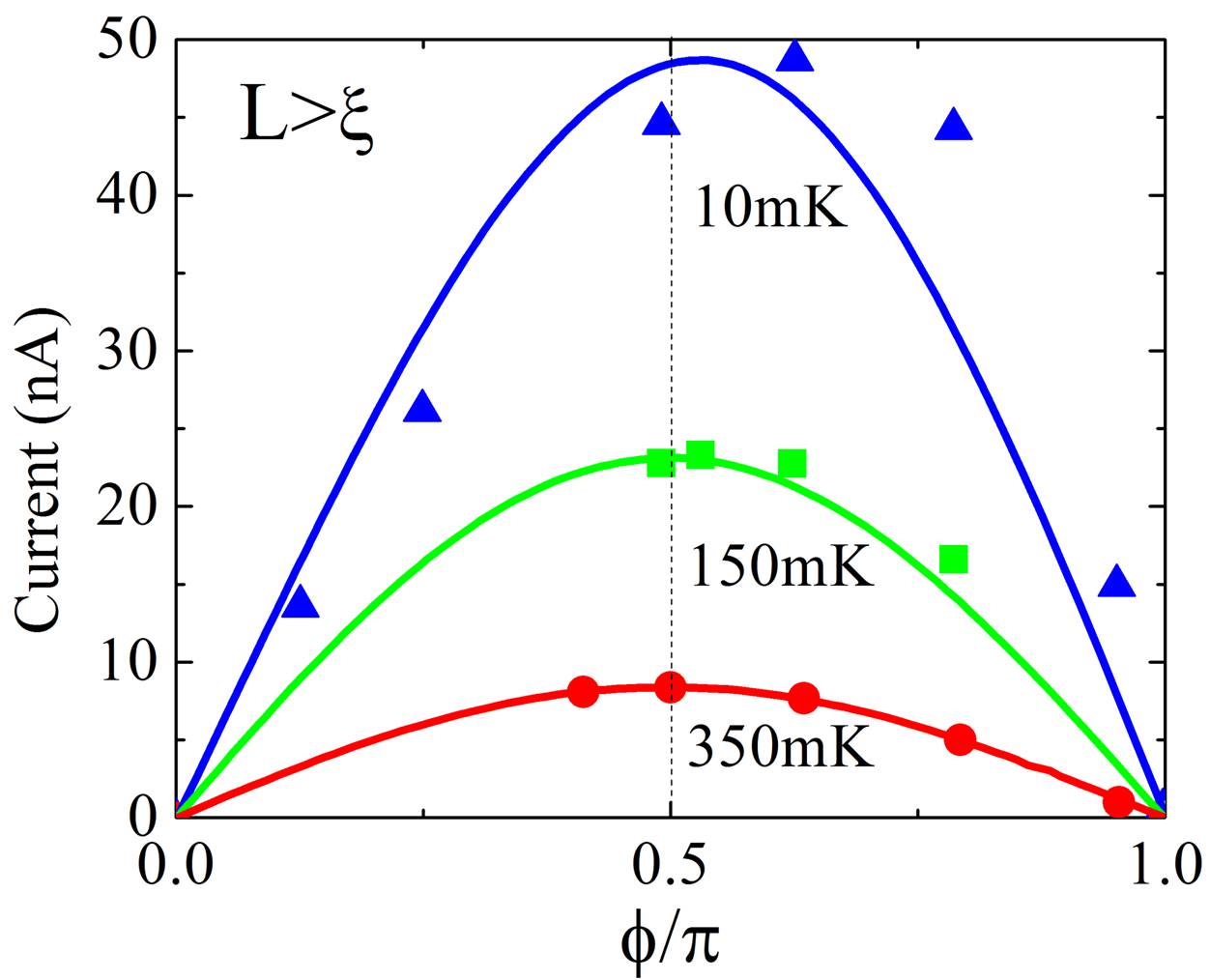}}
\caption{The current phase relation (CPR) at (from top to bottom) 10mK, 150mK, and 350mK in the short junction limit ($L<\xi$, (a)) and the long junction limit ($L>\xi$, (b)). The measured CPR curves are the solid lines.  The dotted lines are theoretical (DBdG method) curves taken from \cite{cserti}.  The scatter points are theoretically calculated (TB-BdG method) in \cite{ABSprimary}.  All theoretical data is scaled to fit the measured curves. At $T = $10 mK, the skewness of the short and long junctions are 0.130 and 0.055, respectively. The TB-BdG method provides the best fits for $L<\xi$.}
\end{figure}

The circuit used to extract the CPR is shown in Fig. 1(a). The Josephson junction is connected in parallel with a fabricated thin film superconducting loop of inductance \textit{L}. A current \textit{I} injected into the circuit divides so that the phases across the junction and the loop are equal (i.e. \begin{math} \phi = 2 \pi (\Phi/\Phi_{o}) \end{math}) where $\Phi$ is the flux in the loop and $\Phi_{o}$ is a flux quantum. The electrodes leading to the junction are intentionally narrow and spread apart to minimize any parasitic geometric inductance that could affect the CPR. The component of current in the loop inductor is measured by coupling the flux in the loop to a commercial dc SQUID via a filamentary superconducting flux transformer.  From the circuit in Fig. 1(a), it can be seen that the current \textit{I$_{J}$} and phase $\phi$ across the junction are given by:

\begin{equation}\label{bladf}
  I_{J}=I-I_{L}=I-\frac{\Phi}L=I-\frac{V_{SQUID}}{V_{\Phi}L}
\end{equation}

\begin{equation}\label{blads}
  \phi=2\pi\frac{V_{SQUID}}{V_{\Phi}L}
\end{equation}
	
where $I_{L}$ is the current through the inductor, and $V_{SQUID}$ is the measured SQUID voltage, with $V_{\Phi}$ the flux transfer function. Noise and thermal drift affecting the flux transformer are removed by measuring the differential SQUID response $\Delta V_{SQUID}$($I$) with a lock-in amplifier. The integration of $\Delta V_{SQUID}$ subsequently produces $V_{SQUID}$($I$). In order to maintain non-hysteretic, ideal behavior in the circuit for CPR extraction, the condition $\beta_{L}=2\pi L I_{c}/\Phi_{o}\leq1$ must be met \cite{tinkham}. While this condition applies exactly for a sinusoidal CPR, it also applies approximately to any CPR that is nearly harmonic. For all samples, $L\sim $2.0 nH, requiring $I_{c}<$ 135 nA in the junctions to extract the CPR. To satisfy this constraint, junction widths were minimized to reduce $I_{C}$. 

Selected CPR curves measured at the Dirac point for Sample A ($L<\xi$) are shown in Fig. 2a.  Clear non-harmonic behavior (forward tilt/skewness) is observed at $T=10$mK and $T=150$mK (similar behavior is observed in Sample B). Also shown are curves and individual points calculated by the Dirac Bogoliubov de Gennes (DBdG) method and the tight binding Bogoliubov de Gennes (TB-BdG) method, respectively \cite{DBdG,ABSprimary} (these methods will be discussed further in the next section). At all temperatures there is a good fit between the measured CPR curves and the CPR curves calculated by the TB-BdG method for $L<\xi$.  The DBdG method provides a worse fit at low temperature, but converges to the measured result above 300mK. Measured CPR curves for Sample C ($L>\xi$) are shown in Fig. 2b. While some forward skewness ($S$ = 0.055) is observed at $T$ = 10 mK, the average skewness is small ($S_{avg}\sim 0.038$, Fig. 7), and is negligible ($S \leq 0.02$) for $T \geq 150$ mK.

\begin{figure}
\subfloat[][]{\includegraphics[width=8.0cm]{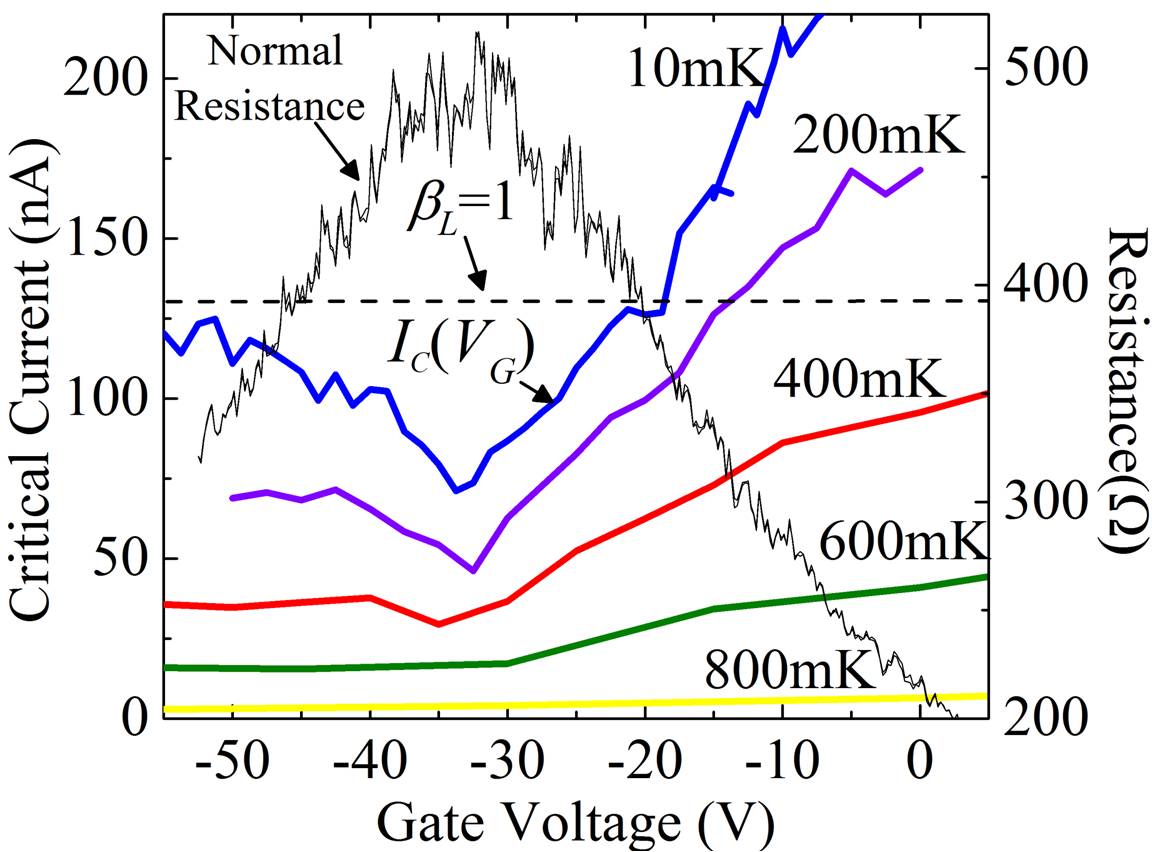}}
\qquad
\subfloat[][]{\includegraphics[width=7.5cm]{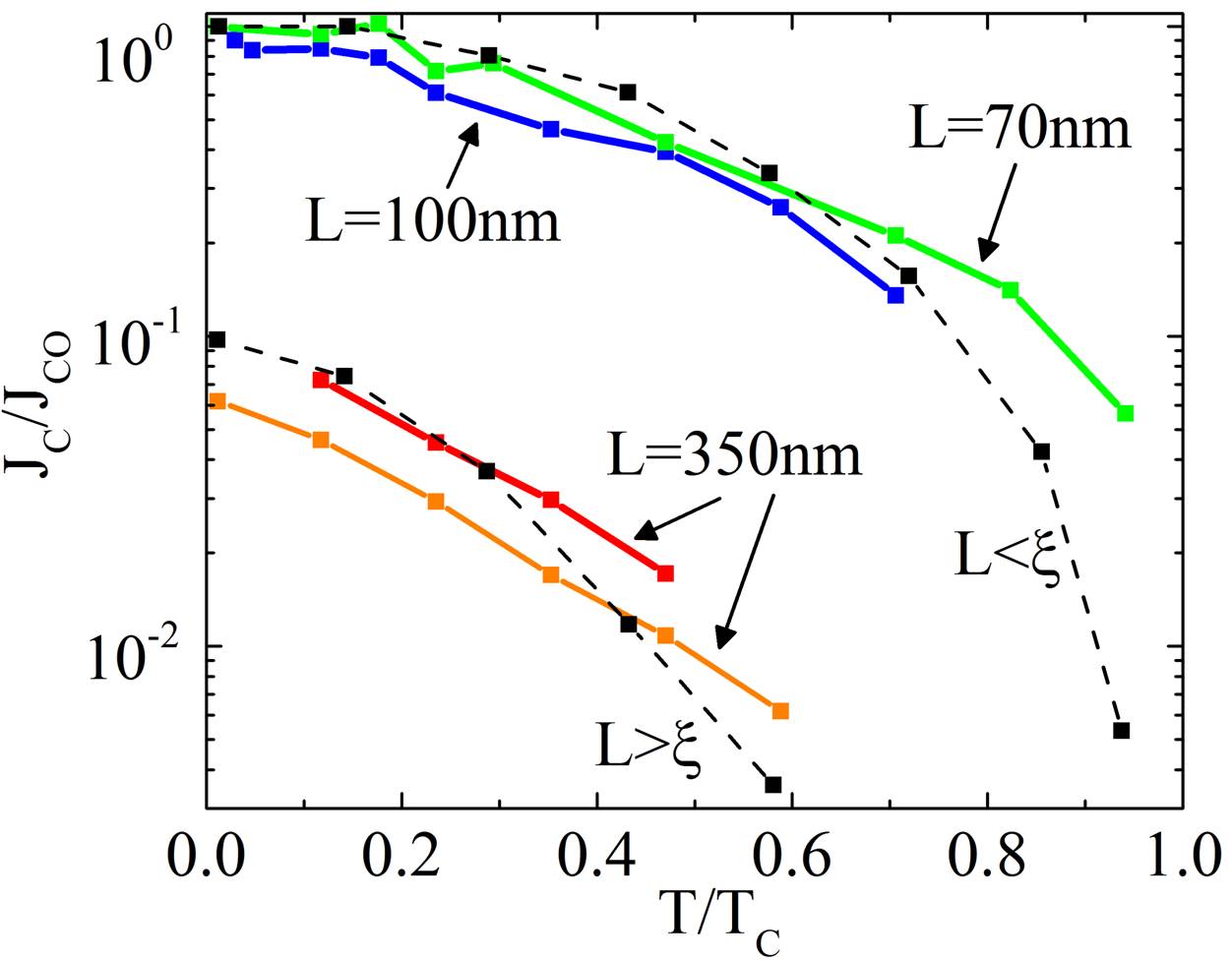}}
\caption{(a) Critical current measurements (left axis) as a function of the back-gate voltage shown at 10mK up to 800mK for Sample A.  The normal resistance at 1.2K (right axis), shown as the solid black line, is also given as a function of gate voltage. The dotted line at $I_{c}=$135nA ($\beta_{L}=1$) marks the maximum critical current at which the CPR can be determined.  (b) Measured critical current as a function of temperature for each sample. Solid lines are measurements, while dashed lines are theoretical fits taken from ref. \cite{ABSprimary}. $J_{co}$=237nA/$\mu$m is taken to be the $J_{c}$ of sample A at 10mK. $J_{c}$ is defined to be $I_{c}/W$.  	}
\end{figure}

\begin{figure}
\subfloat[][]{\includegraphics[width=7.9cm]{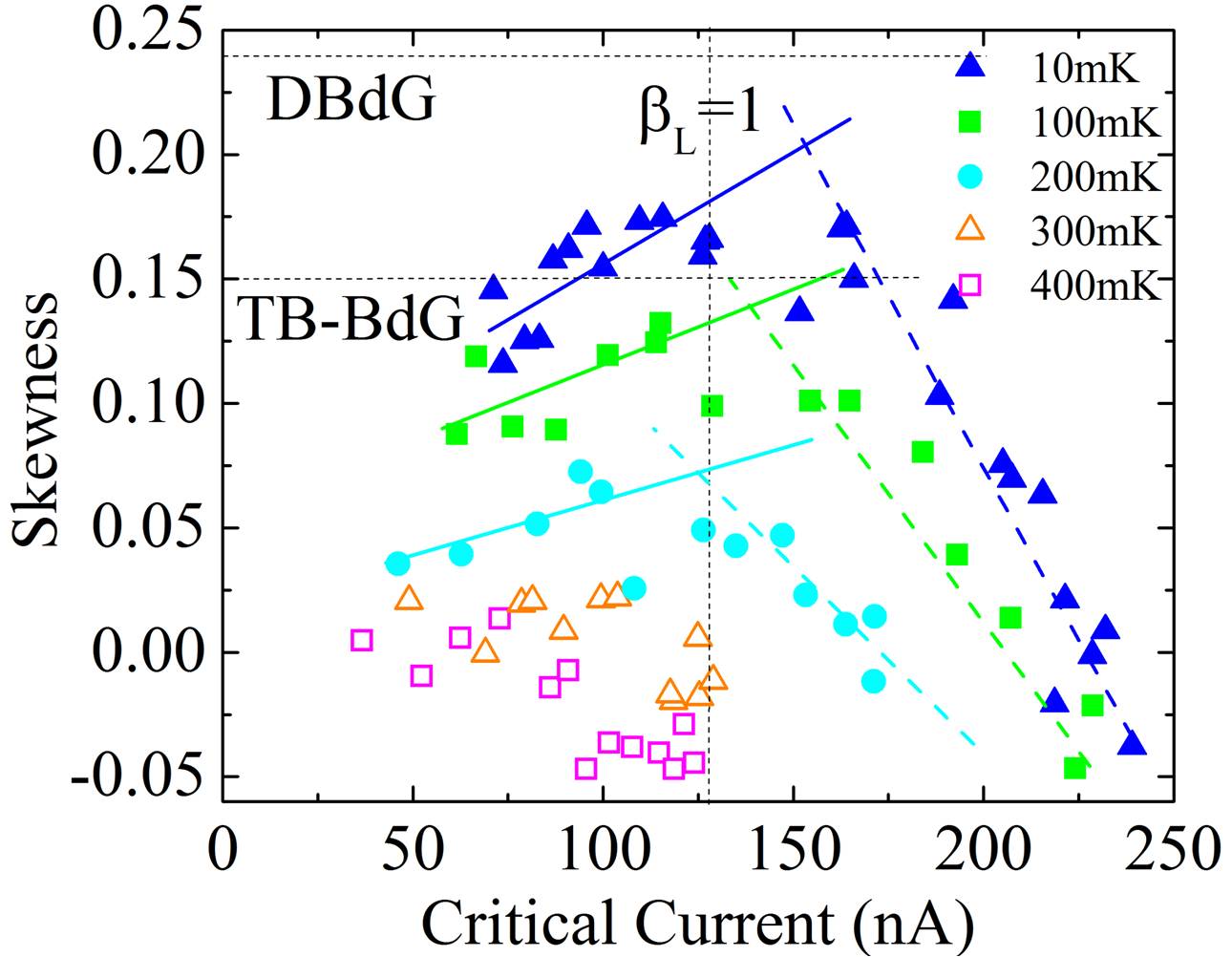}}
\qquad
\subfloat[][]{\includegraphics[width=7.9cm]{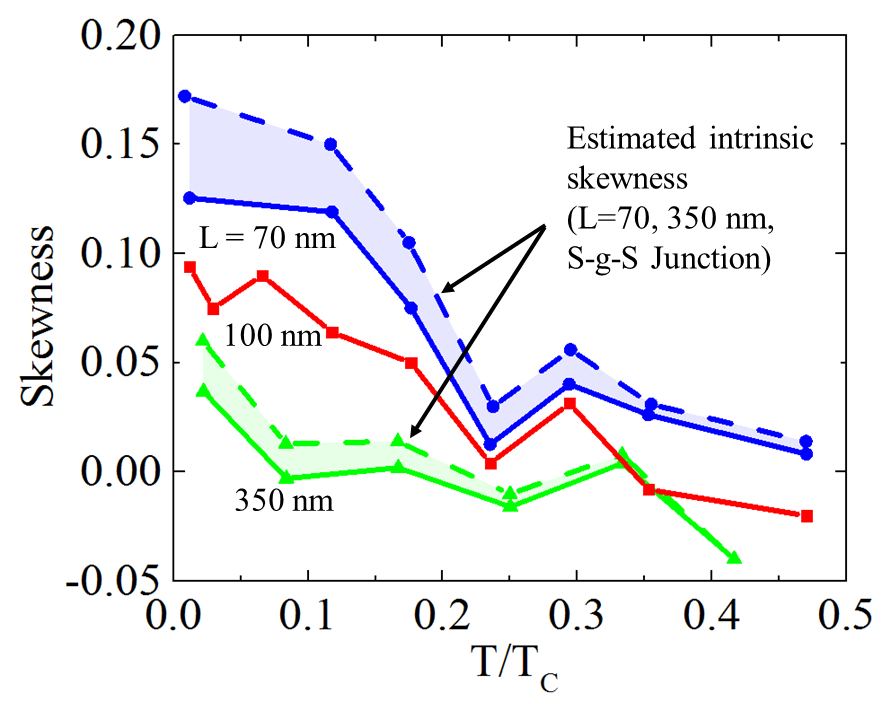}}
\caption{(a) Measured skewness ($S$) of the CPR as a function of critical current ($I_{c}$) for sample A. The horizontal dotted lines represent predictions for the skewness of the CPR at the Dirac point by the DBdG formalism and the TB-BdG formalism.  The vertical dotted line indicates the critical current at which $\beta_{L}=1$.  The solid and dashed lines are guides to the eye for points measured in the non-hysteretic ($\beta_{L}<1$) and hysteretic regimes ($\beta_{L}>1$), respectively. (b) Measured skewness versus temperature for junctions with $L =$ 70, 100, 350 nm (blue, red, green, solid lines) at the Dirac point. The corresponding dashed lines indicate the estimated intrinsic skewness, which is reduced during the measurement by noise-rounding. The upper limit for $L =$ 100 nm (omitted) is similar to that of $L =$ 70 nm.}
\end{figure}

From the CPR measurements, the $I_{c}$ can be extracted as a function of the back-gate voltage $V_{G}$. Fig. 3a displays the $V_{G}$ dependence of $I_{c}$ for sample A, along with the $V_{G}$ dependence of the junction normal resistance $R_{N}$ at low temperature. The Dirac point resides at -35 V (similar to sample B). We consistently observe an asymmetry in $I_{c}$($V_{G}$), common to all samples, that becomes more pronounced at higher temperatures, while $R_{N}$($V_{G}$) appears to be symmetric near the Dirac point: $I_{c}$($V_{G}$) increases sharply on the $n$-doped side of the Dirac point ($V_{G} > -35$V), but increases slowly up to a constant value on the $p$-doped side ($V_{G} < -35$V). This behavior is not well-understood, but has been observed elsewhere \cite{heerche,cho1}. In order to satisfy the constraint $\beta_{L}<1$  and accurately measure the CPR, the $I_{c}$ of the junction should be below the dashed line shown in Fig. 3a. This is achieved at 10mK and 100mK for $V_{G} < 20$V, and is easily satisfied for higher temperatures.  The $J_{c}$($T$) dependence, where $J_{c}=I_{c}/W$, is shown in Fig. 3b for all samples, along with $J_{c}$($T$)  curves calculated by the TB-BdG method in \citep{ABSprimary}. The significant change in $J_{c}$ between junctions with $L<\xi$  and junctions with $L>\xi$ matches the theoretical prediction for short and long junctions.  

\begin{figure}[h]%
\subfloat[][]{\includegraphics[width=4.3cm]{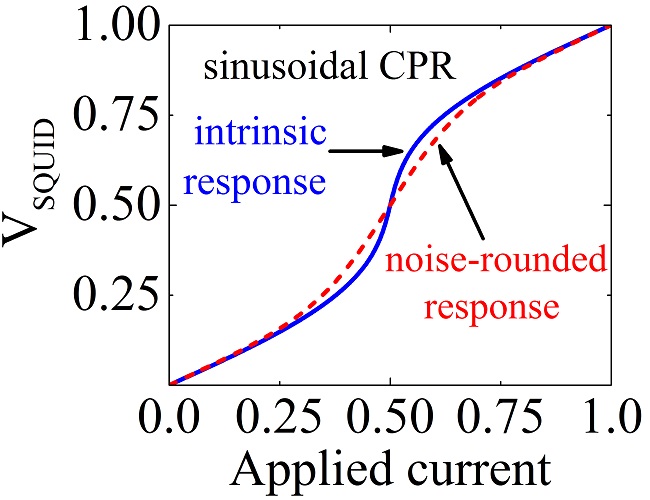}}%
\subfloat[][]{\includegraphics[width=4.3cm]{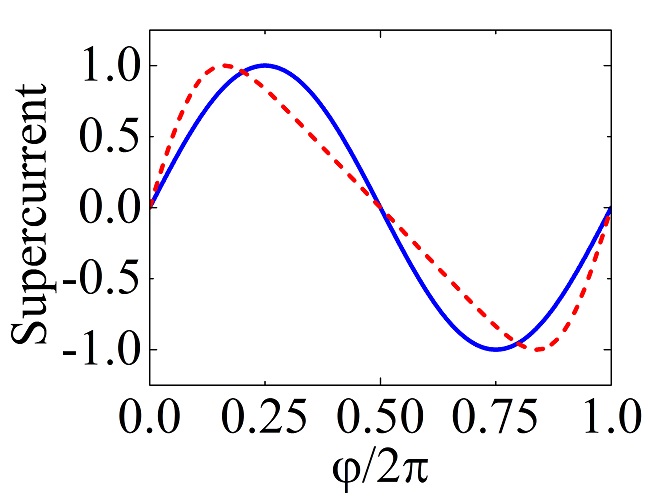}}%
\qquad
\subfloat[][]{\includegraphics[width=4.3cm]{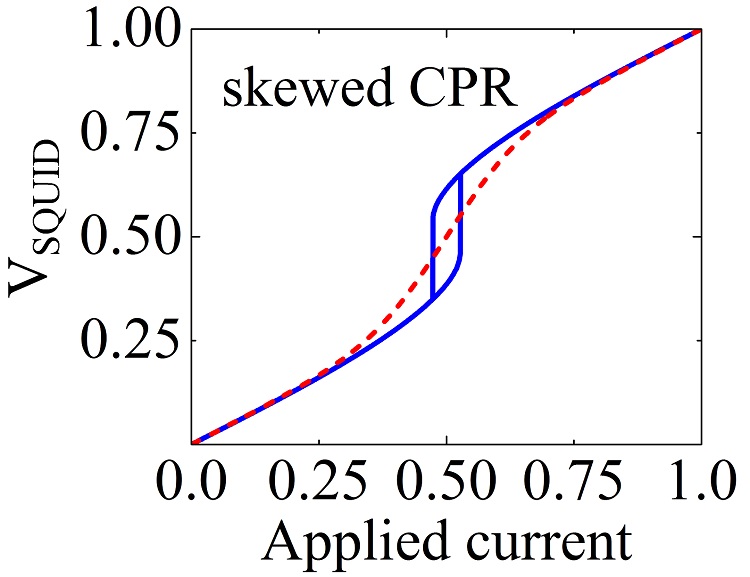}}%
\subfloat[][]{\includegraphics[width=4.3cm]{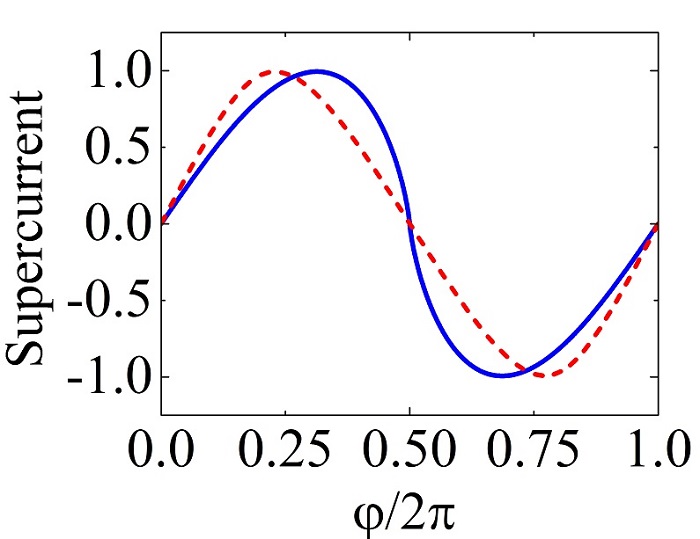}}%

\caption[]{(a) Modeled step response of $V_{SQUID}$ vs applied current for a sin($\phi$) CPR. (b) Extracted CPR from (a). (c) Modeled step response of $V_{SQUID}$ vs applied current for Eq.(4). (d) Extracted CPR from (c). For (a)-(d), the solid line is the intrinsic response and the dashed line is the noise-rounded response.}

\label{fig:cont}
\end{figure}

We may parameterize the skewness (non-harmonic behavior) of the measured CPRs by a variable $S=(2\phi_{max}/\pi)-1$, where $\phi_{max}$ is the position of the maximum of the CPR; $S$ ranges from 0 to 1 as the CPR evolves from a sine wave towards a forward saw-tooth wave.  The $I_{c}$ dependence of $S$ is shown in Fig. 4a for the $n$-doped side of the Dirac cone for Sample A (Sample B shows similar behavior).  Above $I_{c}=135$ nA, $\beta_{L}>1$ and hysteretic switching behavior averaged by noise in the measurement circuit causes the CPR to appear negatively skewed, resulting in a sharp decrease in $S$ vs $I_{c}$. We understand this negative skewness effect as being due to noise-rounding in the CPR measurement.  In particular, the phase interferometer technique depends on measuring the fraction of the applied current flowing through the SQUID loop.  This creates step-like features in $V_{SQUID}$($I$) that, for a sinusoidal CPR, becomes sharp as $\beta_{L}$ approaches 1 and hysteretic for $\beta_{L}>$1.  As a result, noise induced in the circuit can round out these characteristics, distorting the extracted CPR and even removing the hysteresis so that CPR curves can be extracted even in the regime $\beta_{L}>$1.  This effect can be substantial in the graphene junctions because of the large critical currents.
 
\begin{figure}[h]
\includegraphics[width=8.5cm]{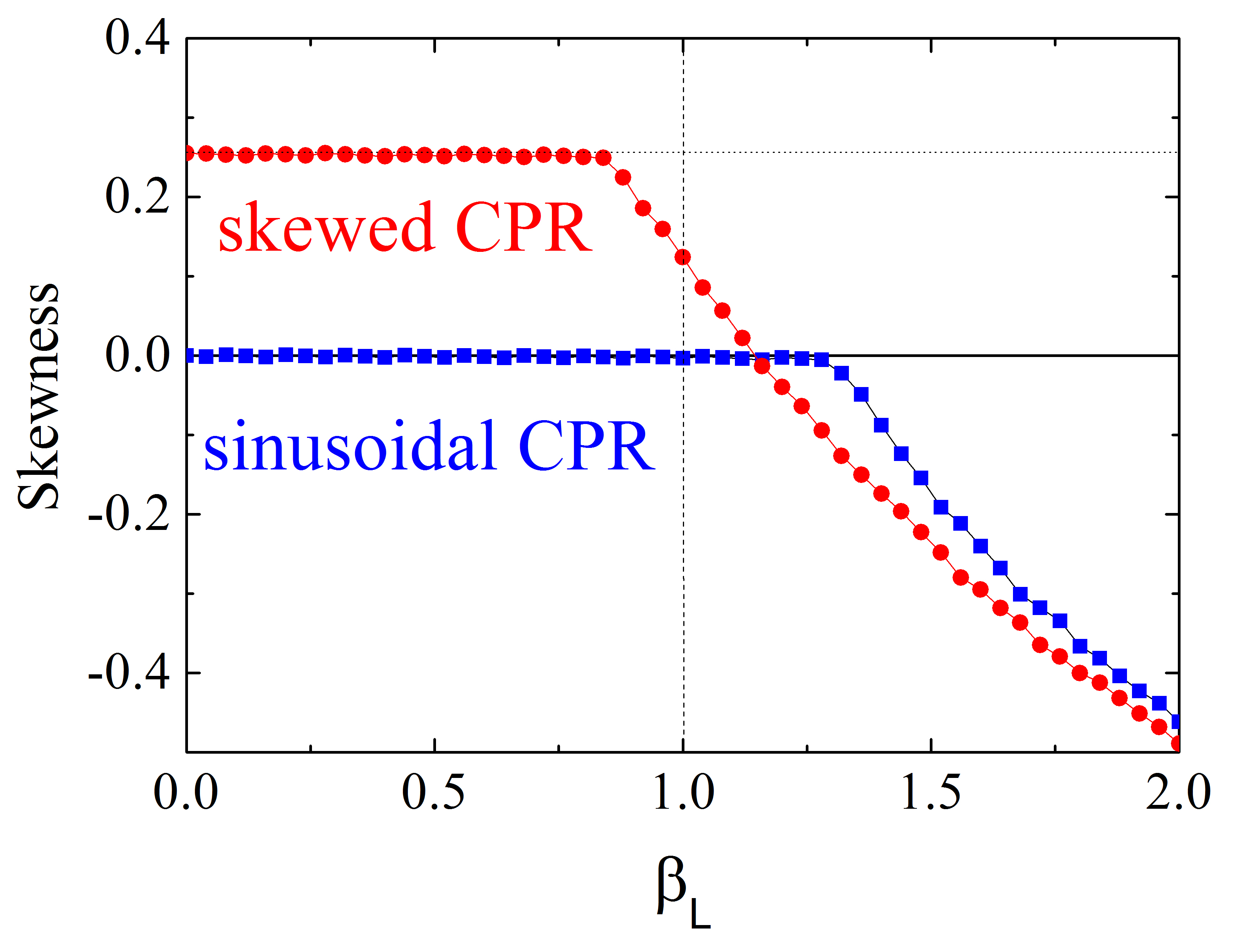}
\caption{Modeled skewness vs $\beta_{L}$ for a sinusoidal CPR and for the skewed CPR given by Eq.(4), demonstrating the effect of noise-rounding on the CPR.}
\end{figure}

In Figures 5a-b, we demonstrate this effect on a pure sin($\phi$) CPR by modeling the SQUID current in the presence of Gaussian noise in the applied current.  The noise preferentially smears out the sharp portions of the steps, resulting in the backwardly-skewed CPR.  This effect is even more pronounced for the intrinsically forward-skewed characteristics that describe the graphene junctions.  For the predicted (and observed) CPR, the SQUID current curves are hysteretic for all values of $\beta_{L}$ so the noise affects the response in all regimes.  An example of how this affects a forward skewed CPR curve is shown in Figures 5c-d.  In Figure 6, we use this noise-rounding model to show the effect of the level of noise on the CPR skewness as a function of $\beta_{L}$.  As expected, the noise-rounding generates a backward skewness that suppresses and then dominates the CPR as $\beta_{L}$ increases.  Comparison with Figure 4a verifies that this mechanism explains this prominent feature of our data.  

However, below $I_{C}=$135 nA, $\beta_{L}<1$  allows for an accurate extraction of the CPR from the measurement. A positive skewness is observed in this regime (Figure 4a) that increases linearly from the Dirac point to $I_{C}=$ 135 nA for temperatures below 300mK. Above 300mK, a much weaker dependence of $S$ on $I_{C}$ is observed. The skewness also exhibits a strong temperature dependence, as shown in Figure 4b. As $T \rightarrow T_{C}$, $S$ approaches 0 for all samples. $S$ increases sharply below 300mK for samples A and B, while it remains close to zero for sample C. $S$ for sample D (not shown) exhibits behavior similar to that of sample C, but is negative below 250mK due to large critical currents ($I_{C}>$135 nA, $\beta_{L}>1$). Note that the measured values of $S$ in Figure 4b are likely an underestimation of the true, intrinsic values, which are reduced by noise rounding. 

We estimate the magnitude of noise in the CPR measurements by fitting our noise model (discussed above) to data in Figure 4a for $T =$ 10 mK and $\beta_{L}>1$. Subsequently, with this noise estimation, we estimate the intrinsic CPR response (before suppression by noise) of the S-g-S junctions (dashed lines, Fig. 4b). The results indicate that the intrinsic $S$ may be up to 35\% higher than the measured values at $T =$ 10 mK. This difference quickly decreases with increasing $T$ (decreasing $\beta_{L}$). Note that the estimated intrinsic $S$ for Sample B ($L =$ 100 nm) is nearly identical to that of Sample A due to larger $I_{C}$ in Sample B. 

Fig. 4 only shows skewness values on the $n$-doped side of the Dirac cone.  The complete set of skewness measurements ($p$-doped and $n$-doped regimes) is shown in Fig. 7 for samples A, B and C at $T$ = 10 mK. Note that the skewness linearly increases for samples A and B on the $n$-doped side of the Dirac cone, while the skewness slightly decreases and then flattens out on the $p$-doped side. Calculations \citep{DBdG} indicate that $S$($V_{G}$) should be symmetrical around the Dirac point, contrary to our observation. Asymmetries in $p$-type and $n$-type conductivities around the Dirac point in graphene have been explained by the difference in scattering cross-section between holes and electrons on charge impurities \cite{novikov, chen}, and also by $p$-$p$ and $p$-$n$ junctions formed by the leads at the graphene interface \cite{huard}. In ref. \cite{huard}, Ti-Al electrodes are reported to produce a higher $n$-type conductivity than $p$-type conductivity, consistent with our observations. This suggests that the contacts are primarily causing the asymmetries observed in our measurements.  
      
\begin{figure}
\includegraphics[width=8.0cm]{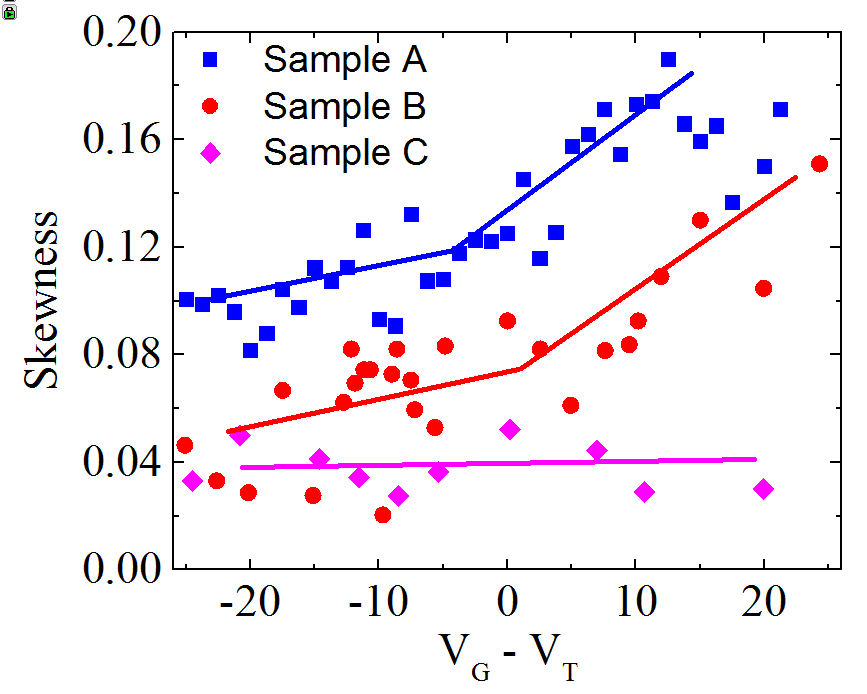}
\caption{Skewness vs. backgate voltage for samples A, B, and C measured at 10 mK. $V_{G}-V_{T}$ is the backgate voltage relative to the Dirac point, where $V_{T}$ is the position of the Dirac point. The solid lines are guides to the eye.}
\end{figure}    
  
In Fig 4, at low $T$, accurate skewness is only measurable up to $n$ = 10$^{12}$ cm$^{-2}$ due to the limitation $\beta_{L}<1$. The carrier density is $n$ = ($V_{G}-$ $V_{Go}$)$C_{ox}$/$q$, where $C_{ox}$ $\sim$ 11.6 nF/cm$^{2}$ for the SiO$_{2}$ substrate, $V_{Go}$ is the voltage at the Dirac point, and $q$ is the elementary charge. However, at higher temperatures ($T$ $\geq$ 300 mK), $I_{C}$ is suppressed such that $\beta_{L}<1$ for $n$ up to 10$^{13}$ cm$^{-2}$. Interestingly, in this high doping regime, the CPR is slightly negative. Both forward and backward skewness have been predicted in graphene at high carrier density (discussed shortly), depending on the influence of the contacts.       
     
The CPR of a ballistic S-g-S junction, at zero temperature and in the short-junction regime, was first predicted theoretically by applying the standard Bogoliubov-de Gennes theory to the Dirac spectrum, also known as the DBdG formalism \cite{DBdG}. $I_{c}$ is carried by Andreev bound states in the junction according to:

\begin{equation}\label{bladz}
  I_{c}=\frac{e\Delta}{h}\sum_{n=0}^{\infty}\frac{T_{n}\mathrm{sin}(\phi)}{\sqrt{1-T_{n}\mathrm{sin}^{2}(\phi/2)}}
\end{equation}

where $\Delta$ is the superconducting energy gap and $T_{n}$ are the transmission coefficients, which are functions of the Andreev bound state wavevectors ($k_{n}$). Large values of $T_{n}$ lead to forward-skewed contributions to $I_{c}(\phi)$. Propagating bound states (real $k_{n}$) that exist for small $n$ have large $T_{n}$ values, and thus contribute the largest amount of skewness. Evanescent bound states (imaginary $k_{n}$) that exist for large $n$ have small $T_{n}$ values, and thus add sinusoidal contributions, reducing the skewness. A closed form solution to eq. 3 at the Dirac point for $W \gg L$ is given by: 

\begin{equation}\label{bladq}
  I_{c}(\phi)=\frac{e\Delta}{h}\frac{2W}{L}\mathrm{cos}(\phi/2)\mathrm{tanh}^{-1}(\mathrm{sin}(\phi/2))
\end{equation}

which has a forward skewness of 0.255. In the DBdG approach, when finite gate voltages are applied to the graphene, the number of propagating bound states increases and dominates the supercurrent. This causes the skewness in the CPR to first increase, then oscillate due to interference effects between the bound states, and then saturate at $S\sim0.42$ with increasing carrier density. The DBdG approach assumes $\Delta$ is fixed at the leads of the superconductor. This boundary condition couples electron and hole transport in the junction, leading to Andreev reflection. We observe skewness values up to 0.17, well below the predictions of DBdG formalism even when accounting for noise-rounding (see Fig. 4b). The DBdG approach has been extended to arbitrary temperatures, but the results still predict skewness values significantly higher than we observe \cite{cserti}. 

The DBdG approach takes into account neither current de-pairing nor proximity effect in the superconducting electrodes, both of which reduce $\Delta$ in the electrodes and have the effect of reducing the skewness. A different approach, referred to as the TB-BdG formalism, incorporates this effect by allowing $\Delta$ to vary in the electrodes as well as in the junction \cite{ABSprimary, ABSsecondary}. The TB-BdG formalism assumes that the superconducting electrodes induce superconductivity in the graphene through an attractive Hubbard pairing potential. An initial guess for $\Delta$($x$) is made, and is re-computed numerically via a self-consistent condition. The results from \cite{ABSprimary} indicate that current de-pairing affects the CPR predominantly in short junctions, while the proximity effect affects the CPR in long junctions. We find that our short junction CPR data at low temperature fits reasonably well to CPR curves calculated by the TB-BdG method (see Fig. 2a) indicating that current de-pairing is likely a significant physical mechanism affecting the junctions. Calculations that include a temperature dependence, but not current de-pairing, are also shown but the curves poorly match the data \cite{cserti}. It is also clear from Fig. 4a that typical measured skewness values at low temperatures are comparable to skewness values predicted by the TB-BdG approach (lower dashed line), while they are below those predicted by the DBdG approach (upper dashed line). 

\begin{figure}
\includegraphics[width=8.0cm]{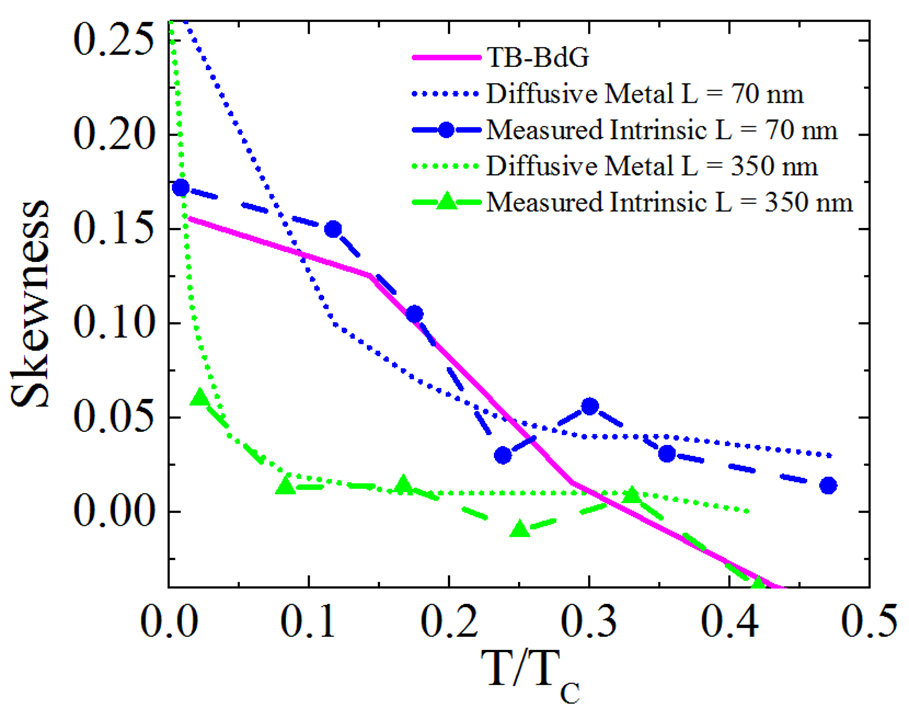}
\caption{Measured intrinsic $S$ vs $T$ for $L$ = 70, 350 nm (blue, green, dashed) compared to calculations with the TB-BdG formalism \cite{ABSprimary} (magenta, solid) and our calculations for a diffusive metal junction with $E_{TH}=$ 3.9, 0.8 $\mu$eV (blue, green, dotted). A reasonable fit is obtained for the short junction with the TB-BdG formalism.}
\end{figure}    


Extractions of the intrinsic $S(T)$ at the Dirac point are compared with calculations from the TB-BdG formalism \cite{ABSprimary} in Fig. 8. The temperature dependence for Sample A, measured near the Dirac point, is in reasonable agreement with the TB-BdG theoretical prediction for $T<0.25T_{C}$, implying that the TB-BdG predictions are reasonable for our data. At the highest temperatures measured ($T$ = 400 mK), where the maximum carrier density is not limited by the requirement $\beta_{L}<$ 1, these devices exhibit a skewness that becomes more negative with increasing $n$, reaching $S$ $\sim$ -0.05 for $n =$ 10$^{13}$ cm$^{-2}$, as indicated in Fig. 4a. The TB-BdG approach \cite{ABSprimary} also predicts a more negative skewness with increasing carrier density in short junctions, reaching up to $S$ $\sim$ -0.5 at $T$/$T_{C}$ = 0.25 when the graphene and metal Fermi levels are aligned. This disparity in skewness most likely results from limitations imposed by our gate oxide, which limits tuning of the graphene Fermi level to +- 0.3 eV \cite{xia}.  The CPR for long junctions can also be calculated using the TB-BdG approach, but our long junction CPR curves (samples C, D) do not match the predicted curves \cite{ABSprimary} (see Fig. 2b)  for low temperature ($T<$ 300mK). However, it is uncertain whether the TB-BdG (or DBdG) approach is applicable to junctions with diffusive ($L$ $\gg$ $l$) transport. 

Although we believe that devices 1 and 2 ($L$ = 70, 100 nm) are operating in a quasi-ballistic regime, we cannot disregard the possible influence of diffusive transport on the junction behavior. The residual doping density ($n_{o}$) of the exfoliated graphene flakes, estimated to be $\sim$5 x 10$^{11}$ cm$^{-2}$, could influence the device to behave more like a typical superconductor-metal-superconductor (SNS) junction. The CPR of a SNS junction is typically forward skewed when thermal fluctuations are less than or equal to the Thouless energy ($E_{TH}$). In a long, diffusive junction, $eI_{C}R_{N}=$ 10.82$E_{TH}$ \cite{Dubos}, where $e$ is the electron charge. Using this expression, we extract $E_{TH}\approx$ 3.9, 0.8 $\mu$eV corresponding to $T= $ 45, 10 mK for $L=$ 70, 350 nm, respectively. For short junctions, $I_{C}R_{N}$ depends solely on the order parameter $\Delta$. Thus, while we do not expect that the above $E_{TH}$ estimate is accurate for $L=$ 70 nm, we use it as a rough approximation. In Figure 8, we simulate the CPR for each $E_{TH}$ based on a model for diffusive junctions \cite{Wilhelm}. For $E_{TH}=$ 3.9 $\mu$eV (corresponding to $L=$70 nm) both our simulation and measurements show an increase in $S$ in the range $T=$ 10-150 mK. However, the diffusive junctions exhibit $S$ up to 0.26 compared to the measured 0.13 - 0.18. By contrast, the result of the TB-BdG formalism \cite{ABSprimary} provides a reasonable fit for our short S-g-S junctions, even when accounting for noise-rounding. For $E_{TH}=$ 0.8 $\mu$eV, the diffusive junction model shows a trend similar to the $L=$350 nm data, suggesting that our long S-g-S junctions might be acting more like disordered metal junctions. Note that some studies \cite{Finkelstein,Bouchiat} experimentally observe $eI_{C}R_{N}\approx$ 0.2$E_{TH}$ for S-g-S junctions, which would increase $E_{TH}$ significantly and (assuming diffusive behavior) yield a constant $S(T)\approx$ 0.26 for the short junctions over the full range $T =$ 0-500 mK; a clear discrepancy from our results.

In conclusion, the CPR curves of graphene Josephson junctions with varying lengths have been measured via a phase-sensitive interferometry technique.  Positive skewness values are reported for short junctions at low temperatures. The measured CPR curves of the short junctions are consistent with those calculated with the TB-BdG formalism. The results suggest that the CPR of a ballistic graphene junction is dominated by a few number of propagating Andreev bound states. This behavior, which is characteristic of Dirac fermions, is similar to typical metallic Josephson junctions. However, to provide a complete picture of the dynamics of the junction, gap suppression due to current de-pairing in the electrodes should be accounted for. Future CPR experiments on junctions with shorter channel lengths ($L<$ 50 nm), preferably with suspended graphene or graphene on a boron nitride substrate \cite{Calado}, should be implemented to obtain fully ballistic transport and confirm our results.
\hspace{10 mm}

We thank Yanjing Lee, Vladimir Orlyanchik, and Martin Stehno for technical assistance. This work was supported by the National Science Foundation under the grant numbers DMR-0906521 and DMR-1411067. Device fabrication was carried out in the Frederick Seitz Materials Research Laboratory at the University of Illinois at Urbana-Champaign. 

\bibliographystyle{prsty} 

\bibliography{myrefs}

\end{document}